\documentclass{article}
\usepackage[utf8]{inputenc}
\usepackage{url}

\usepackage{authblk}

\usepackage{xcolor}

\usepackage{textcomp}
\usepackage{graphicx}
\usepackage{amsmath,amssymb}
\usepackage{xspace}
\newcommand{\etal}{\emph{et al.}\@\xspace}
\newcommand*{\eg}{e.g.\@\xspace}
\newcommand*{\ie}{i.e.\@\xspace}
\newcommand*{\cf}{c.f.\@\xspace}
\usepackage{array}
\newcolumntype{R}[1]{>{\raggedleft\let\newline\\\arraybackslash\hspace{0pt}}m{#1}}
\usepackage{arydshln}
\usepackage[utf8]{inputenc}
\usepackage{mathtools}
\makeatletter
\def\blfootnote{\xdef\@thefnmark{}\@footnotetext}
\makeatother

\title{Initialize globally before acting locally: Enabling Landmark-free 3D US to MRI Registration}
\author[1]{Julia Rackerseder}
\author[2]{Maximilian Baust}
\author[1]{R\"udiger G\"obl}
\author[1,3]{Nassir Navab}
\author[1,4]{Christoph Hennersperger}
\affil[1]{Technische Universität München, Munich, Germany}
\affil[2]{Konica Minolta Laboratory Europe, Munich, Germany}
\affil[3]{Johns Hopkins University, Baltimore, USA}
\affil[4]{Trinity College Dublin, Dublin, Ireland}

\date{\vspace{-5ex}}

\begin{document}

\maketitle

\begin{abstract}
Registration of partial-view 3D US volumes with MRI data is influenced by initialization.
The standard of practice is using extrinsic or intrinsic landmarks, which can be very tedious to obtain. 
To overcome the limitations of registration initialization, we present a novel approach that is based on Euclidean distance maps derived from easily obtainable coarse segmentations.
We evaluate our approach quantitatively on the publicly available RESECT dataset and show that it is robust regarding overlap of target area and initial position.
Furthermore, our method provides initializations that are suitable for state-of-the-art nonlinear, deformable image registration algorithm's capture ranges.
\blfootnote{This project has received funding from the European Union’s Horizon 2020 research and innovation program EDEN2020 under grant agreement No 688279 as well as the GPU grant program from NVIDIA Corporation.}
\blfootnote{This is a pre-print of an article published in the Proceedings of the 21st International Conference on Medical Image Computing and Computer Assisted Interventions (MICCAI), Granada, Spain, September 2018.}
\end{abstract}

\section{Introduction}
\label{sec:introduction}
Image registration, \ie the process of establishing a common reference frame for two or more image data sets, is an important step for a number of medical image computing tasks and computer aided medical procedures.
As noted by Viergerver \etal~\cite{viergever2016survey} in their recent review article on medical image registration, intensity-based approaches are now forming the basis for the vast majority of registration methods, and research in this field focuses almost exclusively on nonlinear image registration.
However, initialization plays a crucial role in convergence of such intensity-based and nonlinear methods.
In case of mono- or multi-modal tomographic registration tasks, such a initialization might be obtained based on the information stored in the header of the respective datasets.
The situation is entirely different for registering 3D ultrasound (US) data, as it lacks a canonical orientation. 
Thus, the registration task is particularly challenging when a common reference frame for 3D US data and Magnetic Resonance Imaging (MRI) data has to be established, because US scans usually depict only a substantially reduced portion of the anatomy.
This is in strong contrast to the capture range of state-of-the-art registration methods, requiring an initial error not greater than 15~mm, as reported recently~\cite{fuerst2014automatic}.

\begin{figure}[tb]
\centering
\includegraphics[height=2.cm]{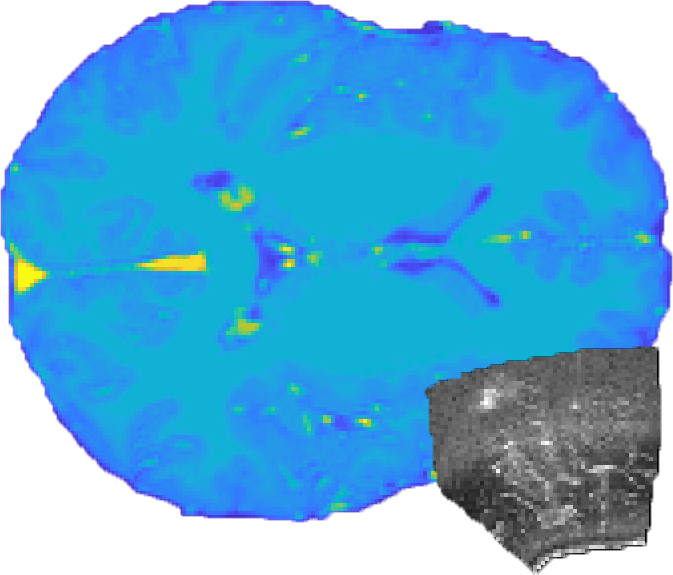}
\hspace{0.4cm}
\includegraphics[height=2cm]{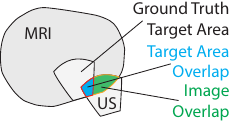}
	\caption{\textbf{Limited overlap in registration.} To initialize registration, a sufficient overlap of images is required. In case of limited overlap landmark selection is challenging. Target Area Overlap is defined as pixels where Target Area and US volume are superimposed (blue), Image Overlap is the part where MRI and US are superimposed before initialization (green + blue).}
	\label{fig:ExplanationOverlap}
\end{figure}

Thus, the application of such nonlinear or \textit{local registration} methods requires a sufficiently close \textit{global initialization}.
If external fiducials are not available or feasible, such an initialization is obtained via the selection of 3D landmarks in common clinical practice.
In view of the aforementioned observations by Viergerver \etal \cite{viergever2016survey}, we argue that the problem of global initialization has received too little attention so far -- particularly for the targeted application of 3D US to MRI registration with limited overlap (see Fig.~\ref{fig:ExplanationOverlap}).  
Although the process of defining a single landmark requires little user interaction (1 click), it depends on profound geometrical understanding of the targeted anatomy as well as the modality-specific appearance.
Particularly in case of 3D US, this process puts a high mental load on the observer, as visual inspection of three dimensional images is difficult due to the lack of predefined orientations as well as the limited volumetric coverage of the anatomy.
While a high precision can be achieved in theory~\cite{xiao2017retrospective}, it is tedious and time consuming.
In practice, this often results in impaired accuracies and high inter-observer variability due to the limited time in daily routine.
Moreover, many works show that the learning curve can be steep when evaluating 3D US, even if the rater had previous training in 2D US~\cite{rodriguez2014learning}.
Contrary to identifying landmarks in 3D, we argue that obtaining coarse segmentations and using them for global initialization is a much more convenient alternative.
The reason is that they can be obtained either with state-of-the-art automatic segmentation techniques, or sophisticated slice-wise and semi-automatic methods.
Furthermore, experts are not required to perform a mental mapping of multiple 3D data sets with partially limited field of view to precisely identify specific and corresponding anatomical landmarks in the data.

We thus propose a novel 
and generally applicable 
initialization procedure based on segmentation-derived distance maps.
We validate this approach on the publicly available REtroSpective Evaluation of Cerebral Tumors (RESECT) dataset~\cite{xiao2017retrospective} 
by using a combination of semi-automatic and automatic segmentation techniques 
and compare it to the global initialization based on landmarks.
\section{Discussion of Related Work}
\label{sec:introduction-relatedwork}
For the nonlinear, deformable registration of 3D US and MRI data, several state-of-the-art methods are available.
They all have in common that initial conditions are stringent in terms of target registration error: for instance, about $15~mm$ are reported by F\"urst \etal \cite{fuerst2014automatic} and below $10~mm$ are reported by Coup\'{e} \etal \cite{coupe20123d}.
In order to obtain an initialization of sufficient quality, three possible methods exist: Usage of external tracking data, landmarks identified in the image data and registration of geometrical entities, \eg rigid registration of segmentations.
If external tracking is not available, such as for retrospective studies, only the latter two strategies are available.
From a clinical point of view, landmark-based initialization appears to be the more widely-used approach, but it requires a sufficient geometrical understanding of the target anatomy and employed imaging modalities as mentioned before.
Reports of inter-observer variation of landmark selection range from $0.33 \pm 0.08~mm$~\cite{xiao2017retrospective} up to $1.6~mm$~\cite{mabee2015reproducibility} even in case of clearly discernible landmarks.
Such observation can also be made for MRI data as demonstrated by Park \etal~\cite{park2015inter}.
As we focus on situations where tracking data is not available, we regard landmark-based initialization as the baseline approach for evaluation, where the aforementioned studies have been used to define a realistic experiment setup, \cf Sec.~\ref{sec:ExperimentsAndResults}.

Segmentation-based registration initialization has been studied in context of prostate fusion biopsy~\cite{fedorov2015open}, where trans-rectal US has to be registered to MRI data.
Both this example and the situation studied in this work (see Fig.~\ref{fig:ExplanationOverlap}) are challenging in terms of limited view of the US volume and the target organ being highly symmetrical, where the global registration of even perfect segmentations would suffer from many ambiguities.

As a consequence, the initialization problem requires further regularization, for which we employ distance transforms which have been shown to be very useful for correspondence estimation~\cite{fedorov2015open,itti1997robust,slavcheva2016sdf}.
Together with an adaptive gradient-based optimization strategy, \cf Sec.~\ref{sec:methods}, we thus are able to satisfy initialization conditions for state-of-the-art deformable registration methods, even in case of very limited views of the US data and coarse semi-automatic US segmentations.
\section{Methods}
\label{sec:methods}
In this section, we derive a novel initialization procedure that only requires low-resolution coarse segmentations to initialize multi-modal deformable 3D US to MRI registration methods.
These segmentations can be easily obtained via coarse annotations or fully automatic methods. 
From these label maps, multi-class distance maps are computed, which are registered simultaneously by optimizing our proposed similarity measure via a gradient-based optimization strategy.
\subsection{Coarse segmentation}
\label{sec:methods-coarseseg}
Let $V_{f}:\Omega_f\rightarrow\mathbb{R}$ denote the fixed and $V_{m}:\Omega_m\rightarrow\mathbb{R}$ the moving volumes defined on their respective domains $\Omega_f,\Omega_m\subset\mathbb{R}^3$.
The first step of our method comprises the creation of $N$ coarse segmentations for both $V_{f}$ and $V_{m}$, \ie we assume two, not necessarily disjoint and complete, partitions of $\Omega_f$ and $\Omega_m$:
\begin{equation}
    \bigcup^N_{\ell=1} \Omega_{f,\ell}\subset \Omega_f\quad\text{and}\quad \bigcup^N_{\ell=1} \Omega_{m,\ell}\subset \Omega_m.
    \label{Equ:Partition}
\end{equation}
The choice of the segmentation algorithm itself depends on targeted anatomy and specific application, but can be automated in most cases.
In Sec.~\ref{sec:ExperimentsAndResults} we evaluate our approach for the application of intra-operative brain imaging, where the US volume takes the role of $V_f$ and the MRI volume takes the role of $V_m$.
\subsection{Initialization Procedure}
\label{sec:methods-init}
Registering the two sets of label masks obtained via segmentation could be formulated as a (pseudo-)mono-modal registration problem for which plenty of classical intensity-based registration techniques are available.
However, this approach would suffer from the following issues:
Firstly, computing the similarity of label maps containing all labels encoded by numerical values would bare the possibility of trading label errors in an unfavorable way: two erroneously registered voxels with a label distance of one would yield the same error as one erroneously registered voxel with label distance two.
Secondly, registering label maps with bad initialization would suffer from low capture range as homogeneous label regions (particularly in case of the background label) would not yield meaningful information for optimization.
In order to overcome these two problems, we propose a similarity measure which computes label-specific distances (taking into account the first problem) and employs distance maps to increase the capture range (solving the latter issue).
We chose distance maps due to their suitability for correspondence estimation, see \cite{itti1997robust,slavcheva2016sdf} for an example.
Therefore, a Euclidean distance transform $\phi$ is applied to each of the $N$ classes individually and the resulting distance maps are denoted by \begin{equation}
    \phi_{f,\ell} = \phi(\chi(\Omega_{f,\ell}))\quad\text{and}\quad \phi_{m,\ell} = \phi(\chi(\Omega_{m,\ell})),
\end{equation}
where $\chi$ denotes the characteristic function applied to the respective set.
This allows us to formulate the initialization task as a minimization problem
  \begin{equation}
     \min_{T\in SE(3)} \sum_{\ell=1}^{N} \int_{\Omega_f} \left| ( \phi_{m,\ell} \circ T)(x) - \phi_{f,\ell}(x) \right| ^p dx,
     \label{Equ:SimMeasure}
  \end{equation}
where $p = 1,2$ and $T\in SE(3)$ denotes the rigid transformation.
As Eq.~\eqref{Equ:SimMeasure} is differentiable, gradient-based optimization techniques can be applied\footnote{In case of $p=1$ a differentiable relaxation can be found.}.
In order to avoid parameter updates from becoming too large and yielding unstable behavior, we employ the following modified gradient descent scheme:
\begin{equation}
    p_{i+1} = p_i - \tau\text{sign}(\delta_i)\min \{ |\delta_i|,p_{\max} \},
\end{equation}
where $p_{i}$ denotes the optimized rotation angle or translation parameter and $\delta_i$ the partial derivative of Eq.~\eqref{Equ:SimMeasure} w.r.t. $p$ at iteration step $i$.
Furthermore, $\tau>0$ is a positive step size parameter and $p_{\max}>0$ regulates the maximum parameter update per iteration.
This way, unstable behavior can be avoided by restricting the maximum parameter update to $\tau p_{\max}$ (measured in radians or mm, respectively).
For $|\delta_i| < p_{\max}$, however, the update scheme corresponds to a regular gradient descent optimization.

The distance maps not only ensure a large capture range, but also cause the cost function in Eq.~\eqref{Equ:SimMeasure} to enjoy favorable properties, as they a more regular than the piecewise constant label maps.
Moreover, from an implementation point of view, it is advisable to employ a foreground mask $\Omega_F$ to restrict the computation of Eq.~\eqref{Equ:SimMeasure} to the target domain $\Omega_F\cap\Omega_f$.
\begin{figure}[t]
    \centering
    \includegraphics[height=3.55cm,trim= 85 285 0 0, clip]{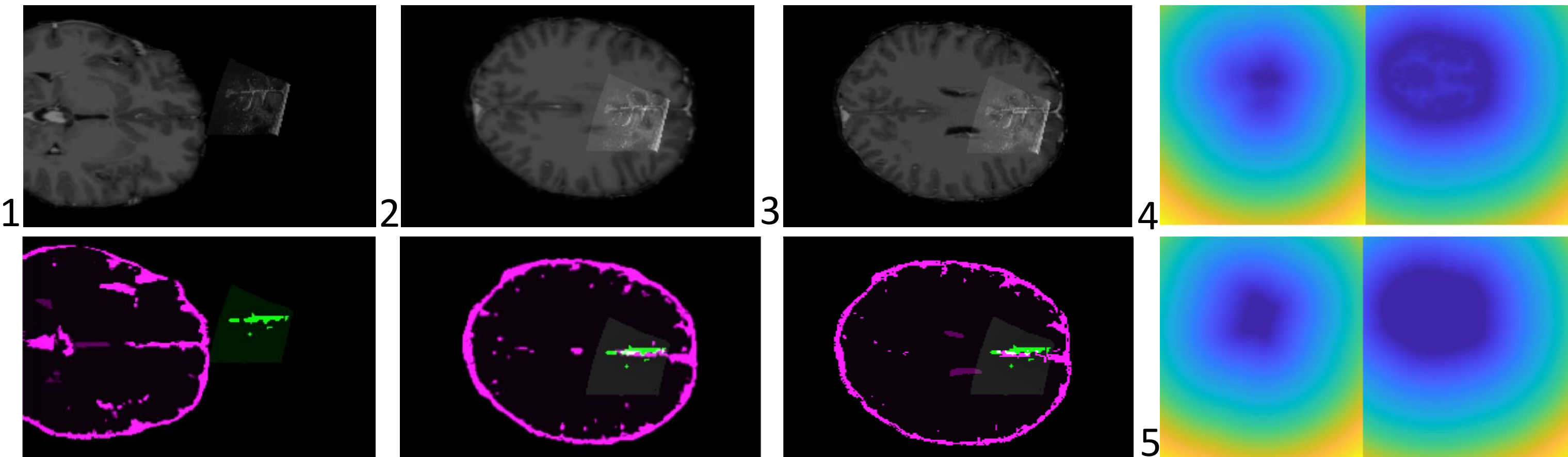}
    \caption{\textbf{Volumes and label map} with (1) $-80~mm$ offset in $x$ direction, $-0.1~rad$ rotation around $\alpha$ and $\beta$ for the MRI (2) after initialization with our method and (3) ground truth provided by RESECT (4) distance map for surface (5) and foreground label.}
    \label{fig:VolumesAndSegs}
\end{figure}
\section{Experiments and Results}
\label{sec:ExperimentsAndResults}
We evaluate our proposed initialization method on the example of the publicly available RESECT dataset~\cite{xiao2017retrospective}. 
It is comprised of imaging data for 23 patients with low-grade gliomas, containing co-registered 3T Gadolinium-enhanced T1w and T2-FLAIR MRI, as well as B-mode ultrasound sweeps from before, during and after tumor resection, reconstructed into 3D volumes.
Retrospectively, up to 17 high accuracy anatomical landmarks were annotated across all three registered US sweeps and between US and MRI volumes for 22 patients.
Only these patients are included in our evaluation.
For easier and faster computation, we downsample all US volumes to match the MRI isotropic resolution of  $1~mm$ in 3D Slicer~\footnote{https://www.slicer.org/}~\cite{Slicer4}.
We mask the foreground in ultrasound and MRI volumes.

With regard to the required coarse registration, the idea is to provide clearly distinguishable and salient labels in both MRI and US, focusing on unique features which are partly visible from any angle the US transducer could be positioned at (see Fig.~\ref{fig:VolumesAndSegs}). 
For brain imaging, included classes are for example (lateral) ventricles, longitudinal fissure and sulci, such as the prominent central and precentral sulcus. 
In other applications, features such as vessel trees, bones, or fasciae could be considered for coarse segmentations.
Due to the penetration depth of the ultrasound in the RESECT dataset, we employ superficial structures, namely sulci, cerebellar tentorium and longitudinal fissure.
Skull stripping and gray-white matter segmentations are automatically performed in FreeSurfer\footnote{http://surfer.nmr.mgh.harvard.edu/fswiki/}~\cite{fischl2002whole}, yielding labels in all MRI datasets that satisfy the characteristics defined above.
For creating the ultrasound label map, we choose the semi-automatic random walk approach~\cite{grady2006random}, where only few pre-labeled pixels are needed.
From the extracted labels, a multi-channel distance map (here, 2 channels: 1 = foreground, 2 = surface) is created for both modalities respectively. The proposed metric (see Eq.~\ref{Equ:SimMeasure}) is estimated and minimized with gradient descent for the distance maps to find the optimal transformation matrix $T$.
We set the step size $\tau$ to 0.5, $p_{max}$ to $0.004~rad$, and $0.5~mm$, keeping updates per step minimal.
\subsection{Evaluation}
\begin{figure}[t]
\centering
\includegraphics[height=3.5cm]{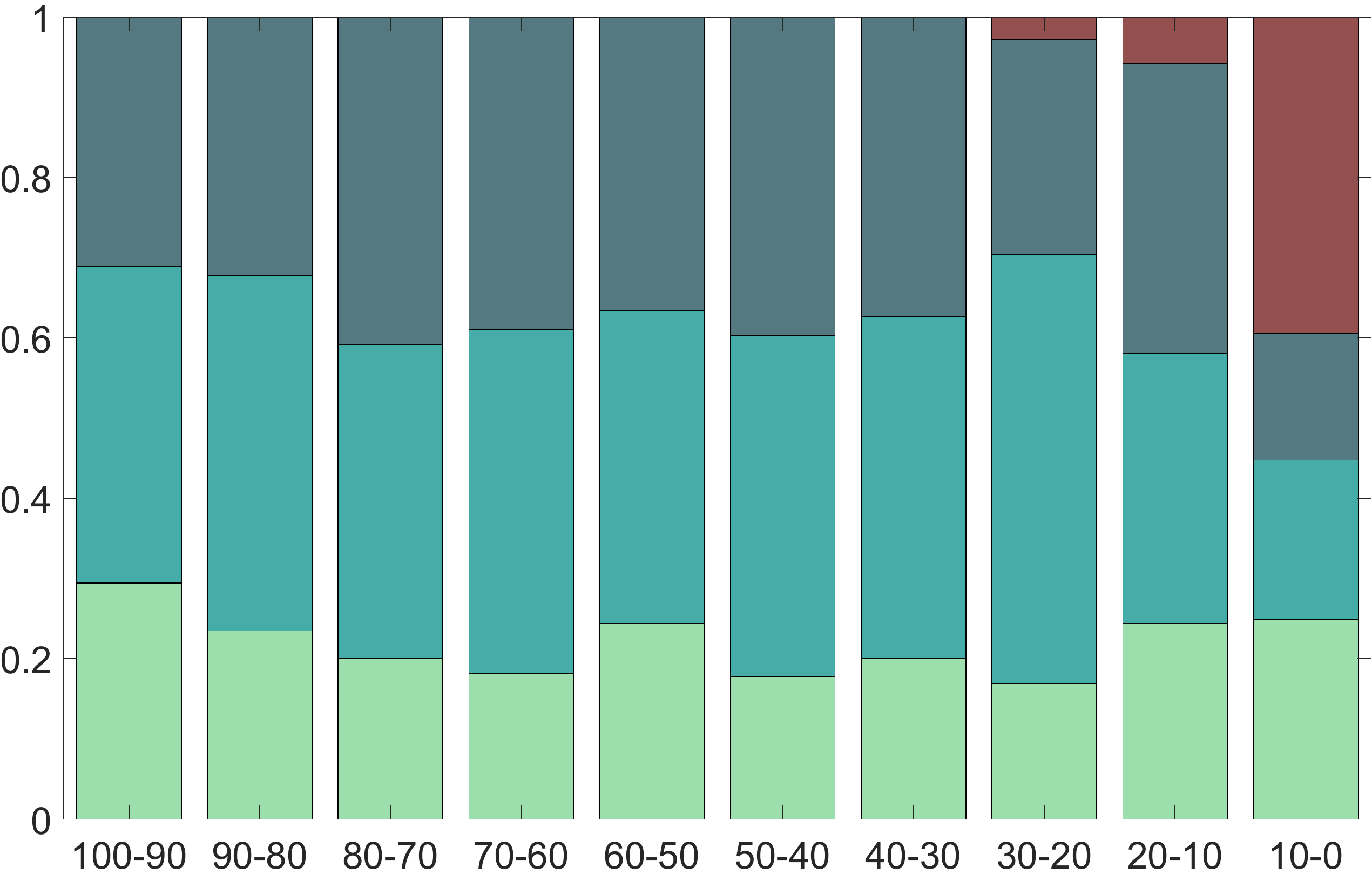}
\includegraphics[height=3.5cm]{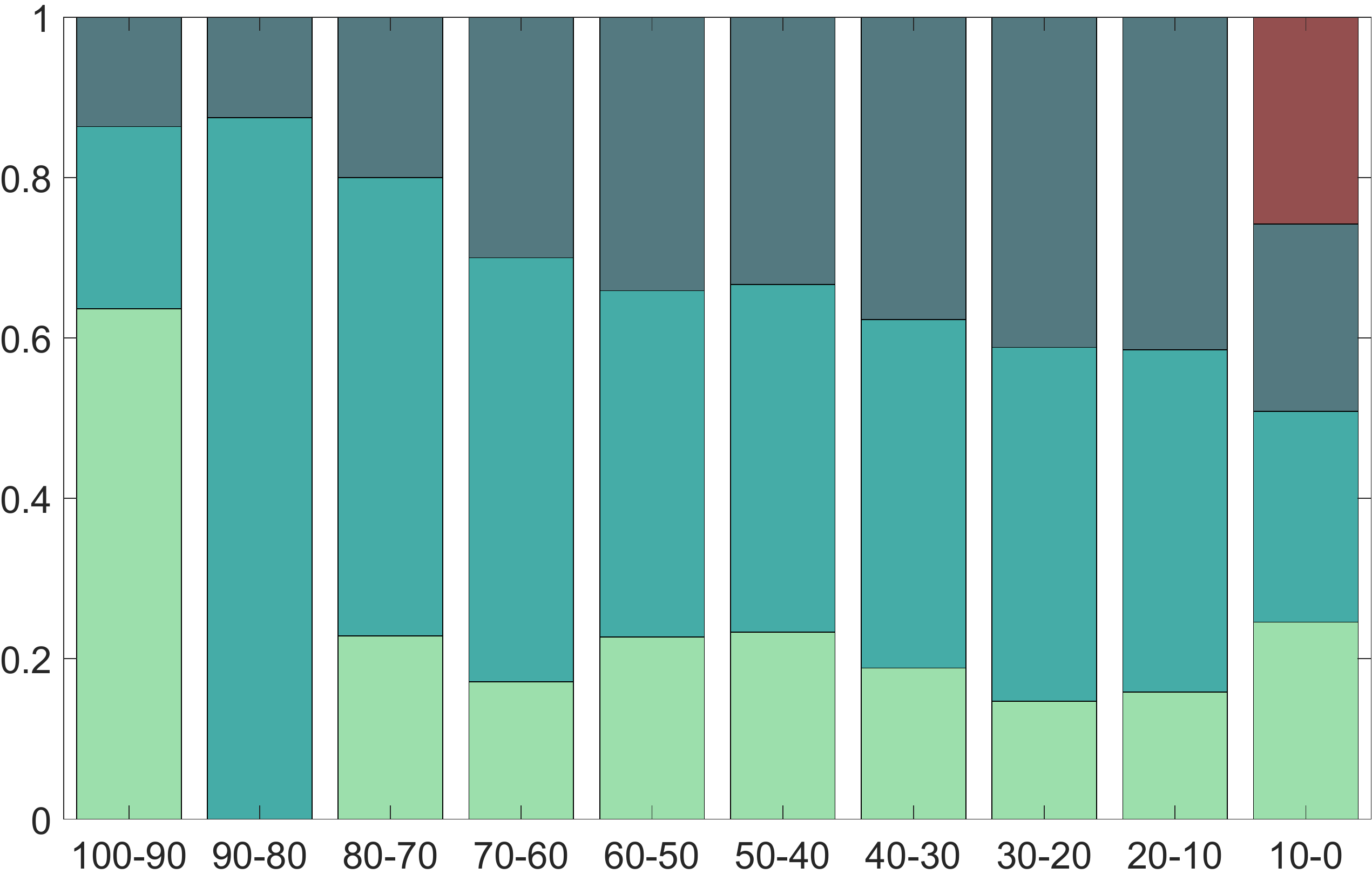}
\includegraphics[height=3.5cm]{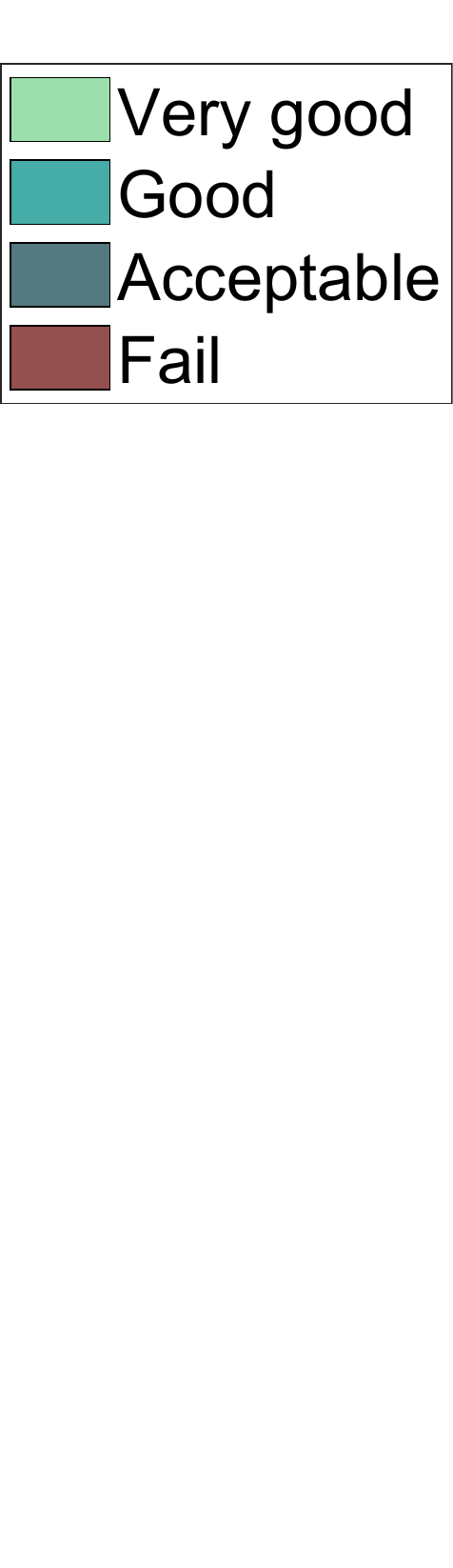}
	\caption{\textbf{Robustness test for decreasing overlap percentage.} These barplots show the fraction of experiments that fall into each quality measure category ($y-axis$) considering the percentage of overlap ($x-axis$) for image overlap (left) and target area overlap right image (right). }
	\label{fig:barplotQualityResults}
\end{figure}
In view of providing a global initialization for following local multi-modal registration, we evaluate the robustness of the proposed initialization, and compare it to manual landmark-annotation as the de-facto standard in practice. 

As a standard error metric for any registration method, the quality of the initialization is evaluated by means of the mean target registration error~\cite{fitzpatrick1998predicting} ($TRE_{mean}$), computed on all landmarks $L$ provided by the RESECT dataset. 

We consider initialization to be a success if the position is within the capture range of state-of-the-art (deformable) registration methods, otherwise we score it as a failure.
With respect to application in neurosurgery, automatic US–MRI registration using the LC$^2$ metric has a capture range of $15~mm$~\cite{fuerst2014automatic}.
Thus we define the following quality criteria:
If $TRE_{mean} \leq 15~mm $ the initialization is considered acceptable, $10-15~mm$ good and $\leq 5~mm$ very good.
\begin{figure}[t]
\centering
\includegraphics[height=3.9cm]{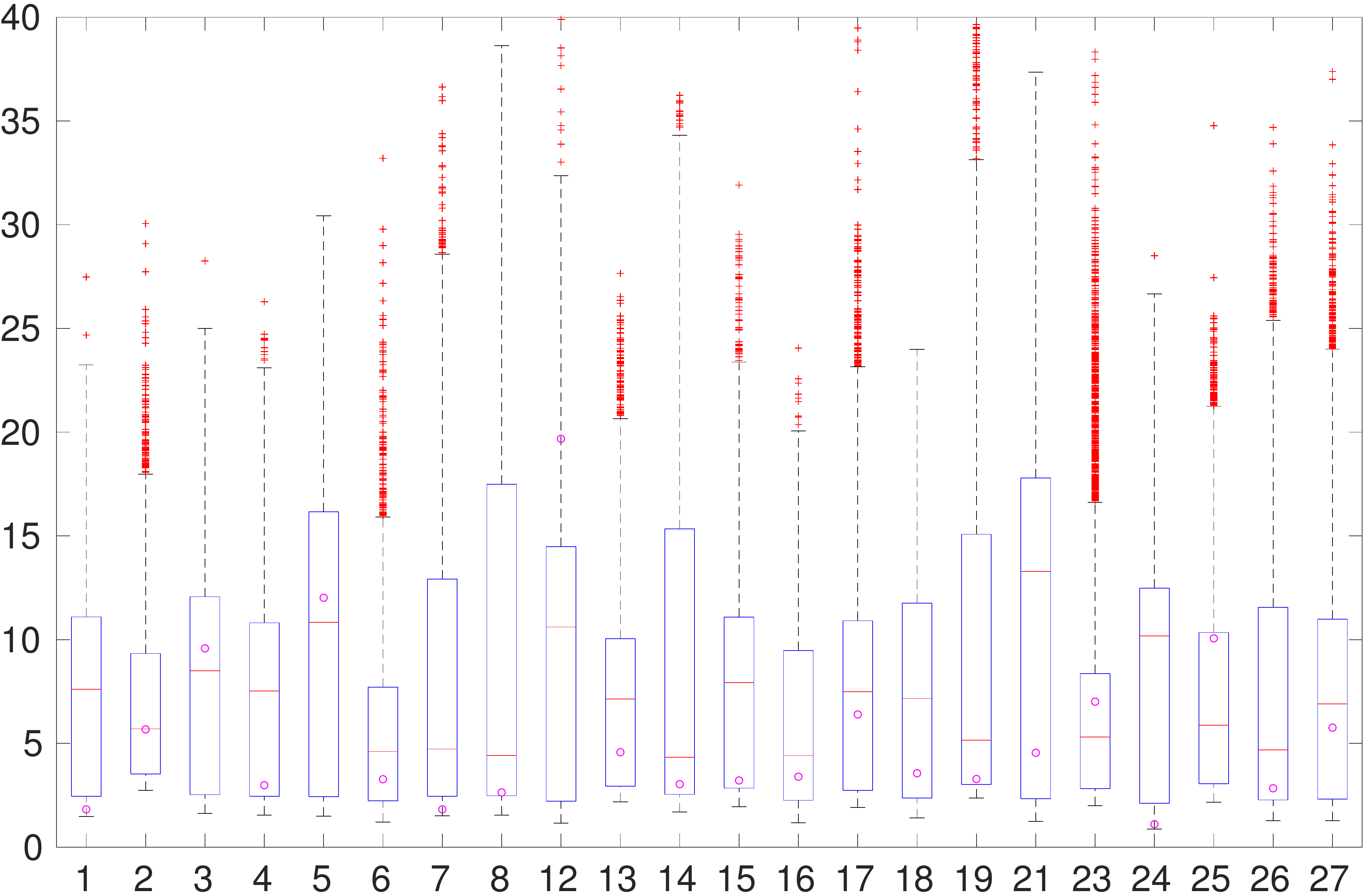}
\includegraphics[height=3.9cm]{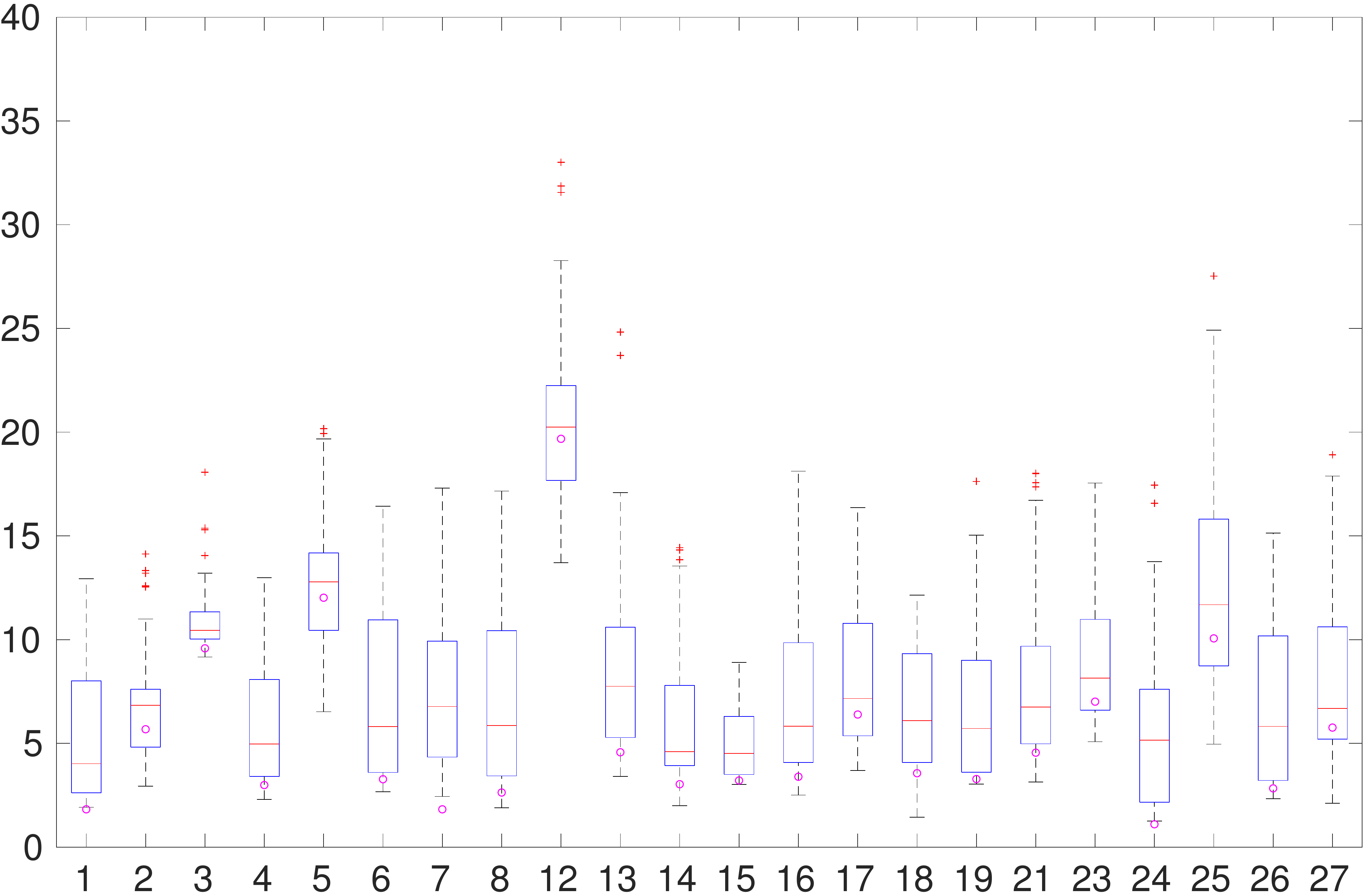}
	\caption{\textbf{Comparison to manual landmark selection.} Shown are errors for all patient TREs for initialization with four random landmarks selected from all available landmark pairs disturbed with Gaussian ($\sigma = 1.5~mm$) noise (left) in comparison to errors for our initialization (right). Circles mark the TRE given by RESECT}
	\label{fig:boxplotLandmarkDisturbance}
\end{figure}

\textbf{Robustness}
In order to test the robustness with regard to target area overlap and image overlap (see Fig.~\ref{fig:ExplanationOverlap}) we conduct convergence tests for increasing translation in x,y,z direction of up to $\pm 200~mm$, as well as rotation around Euler angles $\alpha, \beta, \gamma$ of up to $\pm 0.3~rad$.
In total, this results in 2244 conducted initializations, of which $24.96~\%$ are very good, $32.62~\%$ good, $26.75~\%$ acceptable and $15.64~\%$ fail.
All of the failed cases have below $10~\%$ overlap with the target area. 
Furthermore, all cases with image overlap over $30~\%$ converge with $TRE_{mean} \leq 15~mm$, showing the robustness of the initialization.
Of these, $25.48~\%$ are considered very good, $40.61~\%$ good and $33.91~\%$ acceptable results.
Even $24.94~\%$ of cases with no initial overlap of MRI and US converge with very good results, $19.82~\%$ with good, $15.83~\%$ with acceptable.

\textbf{Comparison to standard in practice}
As discussed in Sec.~\ref{sec:introduction-relatedwork}, the widely used practice is to initialize volumes with non-overlapping positions by manual selection of landmarks.
We simulate this behaviour by randomly choosing 4 landmarks given by the dataset and disturbing them with Gaussian noise with $\sigma = 1.5~mm$, since this is a commonly reported inter-observer variation (see Sec.~\ref{sec:introduction-relatedwork}).
For each patient this is repeated 10,000 times and the $TRE_{mean}$ is calculated on all ground truth landmarks.
Results are visualized in Fig.~\ref{fig:boxplotLandmarkDisturbance} on the left side.
For comparison, on the right side, we show the distribution of $TRE_{mean}$ for our conducted initialization test.

\section{Discussion and Conclusion}
Despite the fact that our results partially show outliers in terms of initialization accuracy, especially the comparison to manual landmark registration, reflects the potentially high inter-operator variability in initialization performance. 
In particular for challenging anatomies, landmark-based registration is demanding for non-experts, because even finding a sufficient number of landmark pairs is often difficult.
In view of applications in practice, it should be noted that many experts are not trained in ultrasound imaging, and thus finding appropriate features can be unclear, also due to quality of US in 3D data.
Even for placing landmarks in MRI high inter-observer variation has been reported \cite{park2015inter}.

Furthermore, the presented initialization is robust with respect to both the target area overlap, as well as the specific image overlap, cf.~Fig \ref{fig:barplotQualityResults}. 
This can be accounted to the specific choice of distance maps in combination with coarse features, providing anatomical context as well as coverage even when the actual volumes do not overlap.
We hope that the proposed method can lead to a simplified clinical routine and more robust results in 3D image registration.

\end{document}